\documentclass[%
 aip,jcp,nofootinbib,
 amsmath,amssymb,
 reprint,
]{revtex4-2}

\usepackage[a-1b]{pdfx}

\usepackage{graphicx}% Include figure files
\usepackage{dcolumn}% Align table columns on decimal point
\usepackage{bm}% bold math
\usepackage[utf8]{inputenc}
\usepackage[T1]{fontenc}
\usepackage{mathptmx}
\usepackage{listings}
\usepackage{rotating} % Rotating table
\usepackage{xcolor}

\definecolor{codegreen}{rgb}{0,0.6,0}
\definecolor{codegray}{rgb}{0.5,0.5,0.5}
\definecolor{codepurple}{rgb}{0.58,0,0.82}
\definecolor{backcolour}{rgb}{0.95,0.95,0.92}
\lstdefinestyle{mystyle}{
  backgroundcolor=\color{backcolour},   commentstyle=\color{codegreen},
  keywordstyle=\color{magenta},
  numberstyle=\tiny\color{codegray},
  stringstyle=\color{codepurple},
  basicstyle=\ttfamily\scriptsize,
  breakatwhitespace=false,         
  breaklines=true,                 
  captionpos=b,                    
  keepspaces=true,                 
  showspaces=false,                
  showstringspaces=false,
  showtabs=false,
  xleftmargin=0.01\textwidth,
  rulecolor=\color[RGB]{200,200,200},
  frame=bt,
  framextopmargin=2pt,
  framexbottommargin=2pt,
  framexleftmargin=10pt,
  tabsize=2
}
\usepackage{color}
\usepackage{dcolumn} % decimal align in tables
\usepackage{bm} % bold math
\usepackage{graphicx}
\usepackage{multirow} % for table cells to span rows
\usepackage{pifont} % for checkmarks
\usepackage{epsfig}
\usepackage{amsmath} % matrix
\usepackage{subfigure}
\usepackage{float}
\usepackage{booktabs}
\usepackage{tabularx}
\usepackage{natbib}
\usepackage{gensymb}
\usepackage{rotating}
\usepackage{threeparttable}
\usepackage{comment}
\usepackage[normalem]{ulem}

\newcommand{\pq}{\texttt{p$^\dagger$q~}}
\newcommand{\np}{\textsc{NumPy}}

\begin{document}
  
\author{Marcus D. Liebenthal$^*$}
\affiliation{
             Department of Chemistry and Biochemistry,
             Florida State University,
             Tallahassee, FL 32306-4390}   
\author{Stephen H. Yuwono$^*$}
\affiliation{
             Department of Chemistry and Biochemistry,
             Florida State University,
             Tallahassee, FL 32306-4390} 

\author{Lauren N. Koulias}
\affiliation{
             Department of Chemistry and Biochemistry,
             Florida State University,
             Tallahassee, FL 32306-4390}   
            
\author{Run R. Li}
\affiliation{
             Department of Chemistry and Biochemistry,
             Florida State University,
             Tallahassee, FL 32306-4390}   

\author{Nicholas C. Rubin}
\affiliation{
             Google Research, 
             Mountain View, CA, USA}
             
\author{A. Eugene DePrince III}
\email{adeprince@fsu.edu}
\affiliation{
             Department of Chemistry and Biochemistry,
             Florida State University,
             Tallahassee, FL 32306-4390}

\title{Automated Quantum Chemistry Code Generation with the \pq Package}

%\begin{document}

%\begin{tocentry}
% \includegraphics[width=3.25in,height=1.75in]{TOC_graphic.tif}
%\end{tocentry}

\begin{abstract}
This article summarizes recent updates to the \pq package, which is a C++ accelerated Python library for generating equations and computer code corresponding to singly-reference many-body quantum chemistry methods such as coupled-cluster (CC) and equation-of-motion (EOM) CC theory. Since 2021, the functionality in \pq  has expanded to include boson operators, coupled fermion-boson operators, unitary cluster operators,  non-particle-conserving EOM operators, spin tracing, multiple single-particle subspaces, and more. Additional developments allow for the generation of C++ and Python code that minimizes floating-point operations via contraction order optimization, subexpression elimination, and the fusion of similar terms.
\end{abstract}

\maketitle

\def\thefootnote{*}\footnotetext{These authors contributed equally to this work.}

% some commands for shorthand

\newcommand{\pqgraph}{\texttt{pq-graph} }

\section{Introduction}

The manual derivation and implementation of many-body quantum chemistry methods can be time consuming and error prone. Fortunately, many common electronic structure methods are expressible in the language of second quantization, which is a convenient formalism that allows one to represent quantum mechanical operators and wave functions in terms of operators that create or destroy particles (creation and annihilation operators, respectively). A benefit of the second quantization formalism is that matrix elements in a many-particle basis that involve creation and annihilation operators are easy to evaluate when the operators are ``normal-ordered'' with respect to a vacuum state. From this point of view, the main technical challenge in deriving equations for many-body methods lies in bringing these operators to normal order, which can be achieved via Wick's theorem,\cite{Wick50_268} diagrammatic techniques,\cite{Cizek66_4256,Bartlett09_book} or by simply rearranging the operators according to their commutation or anti-commutation properties. The algebra of second-quantized operators is amenable to automation, and, as a result, the quantum chemistry community has a long history\cite{Wong73_1,Paldus73_9, Bartlett85_151, Hirao85_8,Bartlett82_1910,Cizek90_831,Schaefer91_1, Bartlett91_387, Paldus96_1210, Schaefer97_7943, Harris99_593, Jeziorski99_1857, Lotrich00_494, Lotrich00_4549, Bartlett00_216, Surjan00_1359, Olsen00_7140, Surjan01_2945, Gauss02_7872, Sherrill04_3374,Kohn09_131101,Kohn10_174117,Kohn10_174118,Gauss13_2639,Neese17_1853,Kohn09_124118,Kohn09_104104,Kohn18_064101,Gauss04_9257,Gauss04_6841,Nooijen05_1,Gauss14_104102,Paldus94_8812, Lotrich01_253,Nooijen02_656,Adamowicz05_024108,Kohn11_204111,Kohn12_204107,Kohn12_131103,Shiozaki15_051103,Shiozaki16_3781,Kohn18_693,Gauss04_9257,Hazra03_4832,Hirata03_9887,Hirata04_51,Sibiryakov06_211,Gauss05_214105,SMITH3,Tew08_201103,Evangelista22_064111,Loos23_085035,DePrince21_e1954709,sequant_github,Scuseria17_184113,Windus06_79,Piecuch21_e1966534,Saue25_2409.06759,Saue25_2504.18516} of developing symbolic algebra tools to streamline the generation of equations for many-body methods and the corresponding executable code (see Ref.~\citenum{Hirata06_2} for a review of such tools).

This paper describes recent developments in the \pq package, which is a C++ accelerated Python library for quantum chemistry code generation.  \pq  was  developed as a tool to facilitate the rapid realization of prototype codes for single-reference electronic structure methods such as many-body perturbation theory (MBPT), coupled-cluster (CC) theory,\cite{Coester58_421,Kuemmel60_477,Cizek66_4256,Cizek69_35,Shavitt72_50,Li99_1,Musial07_291} equation-of-motion (EOM) CC,\cite{Emrich81_379,Bartlett93_7029,Bartlett12_126} or configuration interaction (CI). Since the initial publication describing the library,\cite{DePrince21_e1954709} the functionality in \pq has been expanded to include not only fermion operators relevant to conventional electronic structure theories, but also boson and coupled boson-fermion operators that arise in cavity quantum electrodynamics (QED) generalizations of CC/EOM-CC\cite{Koch20_041043, Manby20_023262, DePrince21_094112, Flick21_9100, Koch21_094113, Flick22_4995, DePrince22_054105, Koch23_031002, DePrince23_5264}  and CI.\cite{Koch21_094113, Foley22_154103,Foley24_1214,Wilson24_094111} The EOM-CC capabilities have also been expanded to include non-particle-conserving excitation operators relevant to ionization potential (IP),\cite{Snijders92_55,Snijders93_15,Gauss94_8938,Krylov11_6028,Krylov12_2726,Wloch05_134113,Wloch06_2854,Piecuch06_234107}  electron attachment (EA),\cite{Bartlett95_3629,Bartlett95_6735,Wloch05_134113,Wloch06_2854,Piecuch06_234107}  and double IP/EA\cite{Stanton03_42,Bartlett97_6812,Piecuch14_868,Piecuch17_3469,Piecuch21_e1966534} forms of EOM-CC theory. Moreover, new active-space specification capabilities allow for the generation of equations and code for active-space formulations of these methods (for example, CC with single and double excitations plus semi-internal triple and/or quadruple excitations,\cite{Adamowicz91_1229,Adamowicz92_3739,Adamowicz93_1875,Adamowicz94_5792,Bartlett99_6103,Piecuch10_2987} {\em i.e.}, CCSDt, CCSDTq, and CCSDtq) or for the core-valence separation (CVS) technique.\cite{Dreuw18_7208} 

Additional enhancements to \pq facilitate the development of production-level implementations of the many-body approaches mentioned above. For example, equations and code generated by the original library were represented within a spin-orbital basis, whereas the current version of \pq can be used to generate spin-traced equations and code for unrestricted CC, EOM-CC, etc. Second, the code generation capabilities of the original library was limited to Python implementations of tensor contractions via calls to \np's \texttt{einsum} (which involved limited floating-point optimization). Since then, we have developed a new module called \pqgraph, which provides enhanced code generation capabilities through graph-based optimizations of the many-body equations.
The \pqgraph module incorporates single-term optimization and subexpression elimination techniques to minimize the number of floating-point operations required for executing the implemented equations. The module also has the capacity to generate optimized code in both Python and C++ (using the syntax of the Tiled-Array library\cite{tiledarray_github}), offering greater flexibility and performance.

The aforementioned capabilities are broadly useful in the context of electronic structure method development, particularly when programmable equations are not available in the literature or no prior codes exist against which results can be checked numerically. For example, this software has been instrumental in the development of QED generalizations of CC and EOM-CC for treating molecules strongly coupled to optical cavity modes. The equations used for the QED-CCSD-1\cite{Koch20_041043} model implemented in Ref.~\citenum{DePrince21_094112} were generated using \pq, as were the C++ codes for the electron-excitation (EE) and electron-attachment (EA) QED-EOM-CC approaches applied in Refs.~\citenum{DePrince23_5264} and \citenum{DePrince22_054105}, respectively, and the QED-CCSD-2 approach\cite{Flick22_4995} applied in Ref.~\citenum{DePrince24_064109}. Along similar lines, new functionality in \pq for unitary CC (UCC) has facilitated the development of families of UCC methods which use different schemes for truncating the BCH expansion of the similiarity-transformed Hamiltonian.\cite{DePrince25_2503.00617} Automatically generated code from \pq can also be used to debug production implementations of literature equations, which was done while developing the relativistic exact two-component (X2C) completely renormalized CC(2,3) and ionization potential (IP) EOM-CCSDT codes  applied in Refs.~\citenum{DePrince24_6521} and \citenum{DePrince25_084110}, respectively.  Additional examples include Refs.~\citenum{DePrince23_044113,DePrince23_054113,Yang23_9177}, which used \pq to generate the elements of the CCSD similarity-transformed Hamiltonian matrix and related equations for time-evolving EOM-CCSD wave functions, and Ref.~\citenum{Rubin24_5068}, which used \pq to extract CI coefficients from cluster amplitudes in the context of tailored and externally corrected CC algorithms in quantum computing applications.

This paper is organized as follows. Section \ref{sec:theory} introduces fermionic and bosonic creation and annihilation operators and the concept of normal order. Section \ref{sec:equation} provides an overview of the functionality in the \pq package that can be used to define operators and wave functions in terms of products of fermionic and bosonic creation and annihilation operators and to bring these products to normal order with respect to a preselected vacuum state. Python code snippets are provided illustrating these concepts, as well as the process of outputting equations and some post-processing steps ({\em e.g.},  introducing spin labels). Section \ref{sec:code} describes how  to generate Python and C++ code corresponding to these equations. Lastly, Sec.~\ref{sec:conclusions} provides some concluding remarks. 

\section{Theory}

\label{sec:theory}

The following conventions are used throughout this work. General electronic spin-orbitals are indexed by the labels $p$, $q$, $r$, $s$, $t$, and $u$. The labels $i$, $j$, $k$, $l$, $m$, and $n$, refer to occupied orbitals. The label $i_\mu$ also refers to an occupied orbital, where $\mu$ is the particular index for that orbital ({\em e.g.}, $i_1$, $i_2$, etc.). The labels $a$, $b$, $c$, $d$, $e$, and $f$, refer to virtual orbitals. The label $a_\mu$ also refers to a virtual orbital, where $\mu$ is, again, a  particular index for that orbital. We use the Einstein summation convention where repeated labels imply summation. 

\subsection{Fermionic and Bosonic Second-Quantized Operators}
  
In many-body quantum chemistry, operators and wave functions are often expressed in terms of products of fermionic or bosonic creation and annihilation operators. The process of evaluating matrix elements in a many-particle basis that involve such quantities is most easily done by bringing the operators to normal order with respect to a chosen vacuum state. The simplification lies in the fact that the expectation value of a normal-ordered set of operators with respect to the vacuum state is zero. The \pq package contains a C++ engine for bringing products of second-quantized operators normal order with respect to a preselected vacuum state by the repeated application of appropriate (anti)commutation relations for the operators.

Fermionic creation ($\hat{a}_p^\dagger$) and annihilation ($\hat{a}_p$) operators obey the following anticommutation relations:
\begin{align}
    \{\hat{a}_p, \hat{a}_q \} = \hat{a}_p \hat{a}_q + \hat{a}_q \hat{a}_p = 0
\end{align}
\begin{align}\{\hat{a}^\dagger_p, \hat{a}^\dagger_q \} = \hat{a}^\dagger_p \hat{a}^\dagger_q + \hat{a}^\dagger_q \hat{a}^\dagger_p = 0
\end{align}
and
\begin{align}
\label{EQN:FERMION_ANTICOMMUTATION}
\{\hat{a}^\dagger_p, \hat{a}_q \} = \hat{a}^\dagger_p \hat{a}_q + \hat{a}_q \hat{a}^\dagger_p = \delta_{pq}
\end{align}
where $\delta_{pq}$ is the Kronecker delta function. In particular, Eq.~\ref{EQN:FERMION_ANTICOMMUTATION} may be used to bring products of fermionic operators to normal order. Similarly, bosonic creation ($\hat{b}^\dagger$) and annihilation ($\hat{b}$) operators obey the commutation relations
\begin{align}
[\hat{b}^\dagger_P, \hat{b}^\dagger_Q ] = \hat{b}^\dagger_P \hat{b}^\dagger_Q - \hat{b}^\dagger_Q \hat{b}^\dagger_P = 0
\end{align}
\begin{align}
[\hat{b}_P, \hat{b}_Q ] &= \hat{b}_P \hat{b}_Q - \hat{b}_Q \hat{b}_P = 0
\end{align}
and
\begin{align}
\label{EQN:BOSON_COMMUTATION}
[\hat{b}_P, \hat{b}^\dagger_Q ] = \hat{b}_P \hat{b}^\dagger_Q - \hat{b}^\dagger_Q \hat{b}_P = \delta_{PQ}
\end{align}
where the labels $P$ and $Q$ refer to boson modes.
Equation~\ref{EQN:BOSON_COMMUTATION} may be used to bring products of bosonic operators to normal order.
While the boson commutator relations above involve multiple boson modes, the \pq package currently only supports a single boson mode. As such, the subscript is suppressed for the remainder of this article.

\subsection{Normal Order}

Let us consider the true vacuum state, $|\rangle = |\rangle_\text{e}|\rangle_\text{b}$, which is a state that contains no particles (neither electrons nor bosons); here, $|\rangle_\text{e}$ and $|\rangle_\text{b}$ represent electron and boson vacuum states, respectively. A string of operators that is normal-ordered with respect to this vacuum state is one where all of the creation operators lie to the left of the annihilation operators.  For example, the following products of fermion operators are all normal-ordered, and their expectation value with respect to the true vacuum state is zero
\begin{align} 
\langle | \hat{a}^\dagger_p | \rangle &= 0 \\
\langle | \hat{a}_q | \rangle &= 0 \\
\langle | \hat{a}^\dagger_p \hat{a}_q | \rangle &= 0 \\
\langle | \hat{a}_p \hat{a}_q | \rangle &= 0 \\
\langle | \hat{a}^\dagger_p \hat{a}^\dagger_q | \rangle &= 0 \\
\text{etc.}\nonumber
\end{align}
Similarly, we have the following cases for boson operators 
\begin{align} 
\langle | \hat{b}^\dagger | \rangle &= 0 \\
\langle | \hat{b} | \rangle &= 0 \\
\langle | \hat{b}^\dagger \hat{b} | \rangle &= 0 \\
\langle | \hat{b} \hat{b} | \rangle &= 0 \\
\langle | \hat{b}^\dagger \hat{b}^\dagger | \rangle &= 0 \\
\text{etc.}\nonumber
\end{align}

Consider an operator that is not normal ordered, $\hat{a}_p \hat{a}^\dagger_q\hat{b}\hat{b}^\dagger$. Bringing this operator to normal order is straightforward, given the relationships in Eqs.~\ref{EQN:FERMION_ANTICOMMUTATION} and \ref{EQN:BOSON_COMMUTATION}. We have
\begin{align}
    \langle | \hat{a}_p \hat{a}^\dagger_q\hat{b}\hat{b}^\dagger | \rangle &= \langle | \delta_{pq}\hat{b}^\dagger\hat{b} | \rangle + \langle | \delta_{pq} | \rangle - \langle | \hat{a}^\dagger_q \hat{a}_p \hat{b}^\dagger\hat{b} | \rangle \nonumber - \langle | \hat{a}^\dagger_q \hat{a}_p | \rangle \\
    &= \delta_{pq}
\end{align}
The only term that does not vanish is the ``fully-contracted'' one that does not include any fermion or boson operators. Now, it is clear how the concept of normal order simplifies the evaluation of integrals over products of second-quantized operators. Once the product is brought to normal order, the only non-zero integrals are the ones involving the fully-contracted terms. 

In single-reference wave function methods like CC, normal order is defined with respect to the Fermi vacuum, which is a single $N$-electron Slater determinant, as opposed to the true vacuum state.
In the \pq package, the Fermi vacuum, $|\Phi_0\rangle$, is defined as 
\begin{align}
    |\Phi_0\rangle = |\Phi_{0,\text{e}}\rangle |\rangle_\text{b}
\end{align}
where $|\Phi_{0,\text{e}}\rangle$ is an $N$-electron Slater determinant and $|\rangle_\text{b}$ is the boson vacuum state.
The $N$-electron state can be built from the electronic vacuum, $|\rangle_\text{e}$, as 
\begin{align}
|\Phi_{0,\text{e}}\rangle = \hat{a}^\dagger_{i_1} \hat{a}^\dagger_{i_2} ... \hat{a}^\dagger_{i_N} |\rangle_\text{e}
\end{align}
For the Fermi vacuum, normal order is chosen such that all operators that annihilate the Fermi vacuum state ($\hat{a}_{a_\mu}$, $\hat{a}^\dagger_{i_\mu}$, or $\hat{b}$) must lie to the right of operators that do not annihilate this state ($\hat{a}^\dagger_{a_\mu}$, $\hat{a}_{i_\mu}$, or $\hat{b}^\dagger$). In this way, any expectation value of normal-ordered operators will vanish, {\em e.g.}, 
\begin{align}
\langle \Phi_0 | \hat{a}^{\dagger}_i |\Phi_0\rangle &= 0 \\
\langle \Phi_0 | \hat{a}_a |\Phi_0\rangle &= 0 \\
\langle \Phi_0 | \hat{a}_j |\Phi_0\rangle &= 0 \\
\langle \Phi_0 | \hat{a}^{\dagger}_b |\Phi_0\rangle &= 0 \\
\langle \Phi_0 | \hat{a}_j \hat{a}^\dagger_i |\Phi_0\rangle &= 0 \\
\langle \Phi_0 | \hat{a}^\dagger_i \hat{a}_a |\Phi_0\rangle &= 0 \\
\langle \Phi_0 | \hat{a}^\dagger_a \hat{a}_i |\Phi_0\rangle &= 0 \\
\langle \Phi_0 | \hat{a}^\dagger_a \hat{a}_b |\Phi_0\rangle &= 0 \\
\text{etc.}\nonumber
\end{align}
As was the case for the true vacuum state, the process of evaluating an expectation value with respect to the Fermi vacuum is simplified by bringing the relevant operator to normal order. The only non-zero terms will be the fully-contracted ones that involve no operators. For more exhaustive discussions of second quantization, including formal definitions of the vacua and creation and annihilation operators, simplification of normal-ordering through Wick's theorem, and diagrammatic methods, the interested reader may consult, for example, Refs.~\citenum{Paldus_NijmegenLectures,Olsen00_book,Wilson07_book,Bartlett09_book} and the references cited therein. Reference \citenum{Stopkowicz24_9572} also provides an overview of the extension of some of these concepts to boson and coupled fermion-boson second-quantized operators.

\section{Equation Generation}

\label{sec:equation}

\subsection{Built-in Operator Types}

\label{sec:operator_types}

Table \ref{tab:operators} provides the symbols and definitions of the second-quantized operator types recognized by \pq, which include bare fermionic and bosonic operators, as well as operators comprised of sums and products thereof. An electronic Hamiltonian can be defined in terms of a general one-body operator (\texttt{h}) and a general antisymmetrized two-body operator (\texttt{g}), or in terms of the Fock operator (\texttt{f}) and the fluctuation potential operator (\texttt{v}). For boson systems and coupled electron-boson systems, \pq supports diagonal boson operators (\texttt{w0}) and products of one-body operators and boson creation (\texttt{d+}) or annihilation (\texttt{d-}) operators. The latter coupled operators could be used to represent the bilinear coupling term in the Pauli-Fierz Hamiltonian,\cite{DePrince23_041301} for example. More complicated Hamiltonians could be constructed from products and sums of any of these operators. Note that the electron orbital labels $p$, $q$, $r$, and $s$ arising in the operators \texttt{h}, \texttt{g}, \texttt{f}, \texttt{v}, \texttt{d+}, and \texttt{d-} are general, meaning that they span both the occupied and virtual spaces. 

For wave functions, \pq supports several operator types that could be used to implement CI, MBPT, CC, and EOM-CC approaches that include up to quadruple electron transitions (\texttt{tn}, \texttt{rn}, and \texttt{ln}, where \texttt{n} = \texttt{1}, \texttt{2}, \texttt{3}, \texttt{4}). Analogous coupled electron-boson operators are also defined for up to quadruple electron transitions plus an arbitrary number (\texttt{m}) of boson creation operators (\texttt{tn,m} and \texttt{rn,m}) or boson annihilation operators (\texttt{ln,m}). In Table \ref{tab:operators}, the left- and right-hand EOM operators are specified by the user for given electronic excitation level, \texttt{n}, but, internally, \pq defines these operators in terms of the number of operators acting in the occupied space ({\em i.e.}, the number of holes created, $n_h$) or the virtual space ({\em i.e.}, the number of particles created, $n_p$). For particle-conserving theories such as excitation-energy (EE) EOM-CC, $n_h = n_p = n$. As is discussed below, \pq also supports non-particle-conserving operators as would arise in the ionization potential (IP), electron attachment (EA), etc., forms of EOM-CC. In such cases, the number of operators acting on the electronic occupied or virtual spaces is adjusted accordingly. 

\begin{sidewaystable*}[!htpb]
\vspace{3.25in}
    \centering
    \caption{Operators supported by \pq.}
    \label{tab:operators}
    \resizebox{\columnwidth}{!}{
    \begin{tabular}{lll}
    \hline\hline
            operator symbol~~~~ & operator definition & operator description\\
        \hline
        \texttt{1}     & $1$ & unit operator \\
        \texttt{a(p)} & $\hat{a}_p$ & electron annihilation operator for orbital $p$ \\
        \texttt{a*(p)} & $\hat{a}^\dagger_p$ & electron creation operator for orbital $p$ \\
        \texttt{b-} & $\hat{b}$ & boson annihilation operator \\
        \texttt{b+} & $\hat{b}^\dagger$ & boson creation operator \\
        \texttt{h}     & $h_{pq} \hat{a}^\dagger_p \hat{a}_q$ & general one-electron operator \\
        \texttt{g}     & $g_{pqrs} \hat{a}^\dagger_p \hat{a}^\dagger_q \hat{a}_s \hat{a}_r$ & general antisymmetrized two-electron operator  \\
        \texttt{f}  & $ f_{pq} \hat{a}^\dagger_p \hat{a}_q$ & Fock operator \\
        \texttt{v}       & $\frac{1}{4} \langle pq||rs \rangle \hat{a}^\dagger_p \hat{a}^\dagger_q \hat{a}_s \hat{a}_r -  \langle pi||qi \rangle \hat{a}^\dagger_p \hat{a}_q$~~~~ & fluctuation potential operator \\
        \texttt{w0} & $w_0 \hat{b}^\dagger \hat{b}$ & diagonal boson operator \\
        \texttt{d+} & $d_{pq} \hat{a}^\dagger_p \hat{a}_q \hat{b}^\dagger$ & product of a one-electron operator and a boson creation operator\\
        \texttt{d-} & $d_{pq} \hat{a}^\dagger_p \hat{a}_q \hat{b} $ & product of a one-electron operator and a boson annihilation operator \\
        \texttt{tn} & $\left(\frac{1}{n!}\right)^2 t^{a_1 \ldots a_n}_{i_1 \ldots i_n} \left ( \prod_{\mu=1}^{n} \hat{a}^\dagger_{a_\mu}\right ) \left ( \prod_{\nu=1}^{n} \hat{a}_{i_{n-\nu+1}} \right ) $ & electron cluster operator (\texttt{n} = \texttt{1}, \texttt{2}, \texttt{3}, \texttt{4}) \\
        \texttt{rn} & $\left(\frac{1}{n_h!}\right)\left(\frac{1}{n_p!}\right) r^{a_1 \ldots a_{n_p}}_{i_{1} \ldots i_{n_h}}  \left ( \prod_{\mu=1}^{n_p} \hat{a}^\dagger_{a_\mu}\right ) \left ( \prod_{\nu=1}^{n_h} \hat{a}_{i_{n_h-\nu+1}} \right )$ & right-hand electron EOM operator (\texttt{n} = \texttt{0}, \texttt{1}, \texttt{2}, \texttt{3}, \texttt{4}) \\
        \texttt{ln} & $\left(\frac{1}{n_h!}\right)\left(\frac{1}{n_p!}\right) l_{a_{1} \ldots a_{n_p}}^{i_1 \ldots i_{n_h}}  \left ( \prod_{\mu=1}^{n_h} \hat{a}^\dagger_{i_\mu}\right ) \left ( \prod_{\nu=1}^{n_p} \hat{a}_{a_{n_p-\nu+1}} \right )$ & left-hand electron EOM operator (\texttt{n} = \texttt{0}, \texttt{1}, \texttt{2}, \texttt{3}, \texttt{4}) \\
        \texttt{tn,m} & $\left(\frac{1}{n!}\right)^2 t^{a_1 \ldots a_n}_{i_1 \ldots i_n, m} \left ( \prod_{\mu=1}^{n} \hat{a}^\dagger_{a_\mu}\right ) \left ( \prod_{\nu=1}^{n} \hat{a}_{i_{n-\nu+1}} \right ) \left ( \hat{b}^\dagger \right )^m $ & electron/photon cluster operator (\texttt{n} = \texttt{0}, \texttt{1}, \texttt{2}, \texttt{3}, \texttt{4}, \texttt{m} $\ge$ \texttt{0}) \\
        \texttt{rn,m} & $\left(\frac{1}{n_h!}\right)\left(\frac{1}{n_p!}\right) r^{a_1 \ldots a_{n_p}}_{i_{1} \ldots i_{n_h}, m}  \left ( \prod_{\mu=1}^{n_p} \hat{a}^\dagger_{a_\mu}\right ) \left ( \prod_{\nu=1}^{n_h} \hat{a}_{i_{n_h-\nu+1}} \right ) \left ( \hat{b}^\dagger\right )^m $ & right-hand electron/photon EOM operator (\texttt{n} = \texttt{0}, \texttt{1}, \texttt{2}, \texttt{3}, \texttt{4}, \texttt{m} $\ge$ \texttt{0}) \\
        \texttt{ln,m} & $\left(\frac{1}{n_h!}\right)\left(\frac{1}{n_p!}\right) l_{a_{1} \ldots a_{n_p}, m}^{i_1 \ldots i_{n_h}}  \left ( \prod_{\mu=1}^{n_h} \hat{a}^\dagger_{i_\mu}\right ) \left ( \prod_{\nu=1}^{n_p} \hat{a}_{a_{n_p-\nu+1}} \right )\left ( \hat{b} \right )^m $ & left-hand electron/photon EOM operator (\texttt{n} = \texttt{0}, \texttt{1}, \texttt{2}, \texttt{3}, \texttt{4}, \texttt{m} $\ge$ 0) \\
        \texttt{e1(p,q)} & $\hat{a}^\dagger_p \hat{a}_q$ & one-electron transition operator \\
        \texttt{e2(p,q,r,s)} & $\hat{a}^\dagger_p \hat{a}^\dagger_q \hat{a}_r \hat{a}_s$ & two-electron transition operator \\
        \texttt{e3(p,q,r,s,t,u)} & $\hat{a}^\dagger_p \hat{a}^\dagger_q \hat{a}^\dagger_r \hat{a}_s \hat{a}_t \hat{a}_u$ & three-electron transition operator \\
        \texttt{e4(p,q,r,s,t,u,v,w)} & $\hat{a}^\dagger_p \hat{a}^\dagger_q \hat{a}^\dagger_r \hat{a}^\dagger_s \hat{a}_t \hat{a}_u \hat{a}_v \hat{a}_w $ & four-electron transition operator \\
        \hline\hline
    \end{tabular}
    }
%\vspace{-0.4cm}
\end{sidewaystable*}

\subsection{Products of Second-Quantized Operators}

\label{SEC:OPERATOR_PRODUCTS}

The following Python code will import the \pq library and initialize a helper object for a specific vacuum state (here, the Fermi vacuum)
\begin{lstlisting}
import pdaggerq
pq = pdaggerq.pq_helper('fermi')
\end{lstlisting}
The helper object contains several functions for defining sums and products of second-quantized operators that arise in many-body quantum chemistry, the most basic of which being
\begin{lstlisting}
pq.add_operator_product(num, ['a','b',...])
\end{lstlisting}
Here, \texttt{num} is a floating-point value, and \texttt{a}, \texttt{b}, etc. represent one of the operators given in Table \ref{tab:operators}. The \texttt{add\_operator\_product} function can be invoked multiple times. Each time it is called, \pq brings this new product of second-quantized operators to normal order and stores the resulting strings of operators internally. In principle, one could build up the normal-ordered operators for an electronic structure method like CI, MBPT, CC, or EOM-CC using only repeated calls to this function. Such an approach could be tedious or error prone, so the \pq helper object also includes functions corresponding to other standard operations that appear in these methods. 

Commutators, nested commutators, and anticommutators of products of operators can be defined using
\begin{lstlisting}
pq.add_commutator(num, o1, o2)
pq.add_double_commutator(num, o1, o2, o3)
pq.add_triple_commutator(num, o1, o2, o3, o4)
pq.add_quadruple_commutator(num, o1, o2, o3, o4, o5)
pq.add_anti_commutator(num, o1, o2)
\end{lstlisting}
Here, \texttt{o1}, etc. refer to lists of operators defined in Table \ref{tab:operators}. Each of these lists is interpreted as a product of operators. For example, 
\begin{lstlisting}
pq.add_double_commutator(0.5, ['a','b'],['c'],['d','e'])
\end{lstlisting}
would correspond to the mathematical expression $\frac{1}{2}\left [\left [\hat{a} \hat{b}, \hat{c}\right ], \hat{d} \hat{e}\right ]$. The similarity transformation of a product of operators can be defined using
\begin{lstlisting}
pq.add_st_operator(num, ['a','b',...],['c','d',...])
\end{lstlisting}
where the first list of strings defines a product of operators to be transformed, and the second list of strings represents a sum of operators that defines the transformation. Internally, the similarity transformation is represented using the Baker-Campbell-Hausdorff (BCH) expansion, so, for this example, we would have
\begin{align}
\label{EQN:BCH}
   & \text{exp}\left (-\hat{c}-\hat{d}-...\right ) \left(\hat{a}\hat{b}...\right)\text{exp}\left (\hat{c}+\hat{d}+...\right ) \nonumber \\
    & = \hat{a}\hat{b}... + \left [ \hat{a}\hat{b}..., \hat{c} + \hat{d} + ... \right ]\nonumber \\
    &+ \frac{1}{2!} \left [ \left [ \hat{a}\hat{b}..., \hat{c} + \hat{d} + ...\right ], \hat{c} + \hat{d} + ...\right ] \nonumber \\
    &+ \frac{1}{3!} \left [ \left [ \left [ \hat{a}\hat{b}..., \hat{c} + \hat{d} + ...\right ], \hat{c} + \hat{d} + ...\right ], \hat{c} + \hat{d} + ...\right ] \nonumber \\
    & + ...
\end{align}
Note that \pq makes two assumptions in the \texttt{add\_st\_operator} function. First, it is assumed that the BCH expansion truncates after four nested commutators, which should be the case for most use cases in quantum chemistry, with some exceptions ({\em e.g.}, unitary CC [UCC] theory\cite{Noga89_133,Bartlett89_359}). Second, it is assumed that the operators that define the transformation ($\hat{c}$ and $\hat{d}$ in Eq.~\ref{EQN:BCH}) commute, which greatly reduces the computational effort required to bring the resulting operators to normal order. This assumption is valid in conventional CC theory but not in UCC theory. For use cases involving non-commuting operators, the user may pass an optional argument to the \texttt{add\_st\_operator} function that lifts this assumption (\texttt{do\_operators\_commute = False}).

\subsection{Bra and Ket States}

As mentioned above, the \texttt{add\_operator\_product} function alone could be used to build up strings of normal-ordered operators relevant to many common quantum chemistry methods. The commutator, anticommutator, and similarity transformation functions introduced in the previous section simplifies the process. The specification of general expressions involving second-quantized operators can be further streamlined with functions that define custom bra and ket states. For example, in CI or EOM-CC, one might wish to evaluate a right-hand $\sigma$-vector, which is the action of the (similarity-transformed) Hamiltonian on a ket state spanning some many-particle Hilbert space. As a specific example, consider the doubles part of the right-hand $\sigma$-vector in EE-EOM-CC with single and double excitations (EE-EOM-CCSD), 
\begin{align}
\label{eqn:eomcc_sigma2}
    \sigma^{ab}_{ij} = \langle \Phi_0 | \hat{a}^\dagger_i \hat{a}^\dagger_j \hat{a}_b \hat{a}_a 
    \bar{H} \left ( \hat{R}_0 + \hat{R}_1 + \hat{R}_2 \right ) | \Phi_0 \rangle
\end{align}
with 
\begin{align}
    \bar{H} = \text{exp}\left(-\hat{T}_1 - \hat{T}_2\right ) \hat{H} \text{exp}\left(\hat{T}_1 + \hat{T}_2\right )
\end{align}
Here, $\hat{H} = \hat{f} + \hat{v}$, where $\hat{f}$ and $\hat{v}$ are the Fock and fluctuation potential operators, respectively, and the operators $\hat{T}_1$, $\hat{T}_2$, $\hat{R}_0$, $\hat{R}_1$, and $\hat{R}_2$ correspond to \texttt{tn} and \texttt{rn} in Table \ref{tab:operators}, with appropriate choices for \texttt{n}. Equation \ref{eqn:eomcc_sigma2} can be evaluated in \pq using the following code
\begin{lstlisting}
pq.set_left_operators([['a*(i)','a*(j)','a(b)','a(a)']])
pq.set_right_operators([['r0'],['r1'],['r2']])
pq.add_st_operator(1.0,['f'],['t1','t2'])
pq.add_st_operator(1.0,['v'],['t1','t2'])
\end{lstlisting}
The \texttt{set\_left/right\_operators} functions take as an argument a list of lists of strings, where the inner lists represent products of operators, and the outer list represents a sum of these products. 

The preceding example is specific to a particle-conserving theory where the \texttt{rn} operator contains an equal number of electronic creation and annihilation operators, but it is easily generalizable to non-particle-conserving theories. \pq supports non-particle-conserving left-hand (\texttt{ln} and \texttt{ln,m}) and right-hand (\texttt{rn} and \texttt{rn,m}) EOM operators that result in the addition/removal of up to two electrons to/from the bra or ket states, respectively. Such operators could correspond to the IP, EA, double IP (DIP), and double EA (DEA) flavors of EOM-CC theory. As an example,  consider the 2-hole-1-particle part of the right-hand $\sigma$-vector in IP-EOM-CCCSD
\begin{align}
\label{eqn:ip_eomcc_sigma2}
    \sigma^{a}_{ij} = \langle \Phi_0 | \hat{a}^\dagger_i \hat{a}^\dagger_j \hat{a}_a 
    \bar{H} \left ( \hat{R}_1 + \hat{R}_2 \right ) | \Phi_0 \rangle
\end{align}
again, with 
\begin{align}
    \bar{H} = \text{exp}\left(-\hat{T}_1 - \hat{T}_2\right ) \hat{H} \text{exp}\left(\hat{T}_1 + \hat{T}_2\right )
\end{align}
The corresponding code for this expression is
\begin{lstlisting}
pq.set_right_operators_type('IP')
pq.set_left_operators([['a*(i)','a*(j)','a(a)']])
pq.set_right_operators([['r1'],['r2']])
pq.add_st_operator(1.0,['f'],['t1','t2'])
pq.add_st_operator(1.0,['v'],['t1','t2'])
\end{lstlisting}
For the \texttt{rn} and \texttt{rn,m} operators in Table \ref{tab:operators}, the \texttt{set\_right\_operators\_type} function adjusts the number of electron annihilation operators acting on occupied orbitals (creating holes, $n_h$) and annihilation operators acting on virtual orbitals (creating particles, $n_p$) for a given flavor of EOM-CC. In the case of IP-EOM-CC, $n_h = n$ and $n_p = n - 1$. A similar function (\texttt{set\_left\_operators\_type}) can be used to adjust $n_h$ and $n_p$ for \texttt{ln} and \texttt{ln,m}. For both the left- and right-hand EOM operators, the default operator type is \texttt{EE}, which corresponds to particle-conserving operators. 
Table \ref{tab:non_particle_conserving} outlines how $n_h$ and $n_p$ are defined for different EOM operator types.

\begin{table}[!htpb]
    \centering
    \caption{The number of creation/annihilation operators acting on the occupied ($n_h$) and virtual ($n_p$) orbital spaces for a given flavor of EOM-CC. }
    \label{tab:non_particle_conserving}
    \begin{tabular}{lll}
    \hline\hline
            operator type~~~~ & $n_h$ & $n_p$ \\
        \hline
\texttt{EE}  & $n$ & $n$ \\
\texttt{IP}  & $n$ & $n-1$ \\
\texttt{DIP} & $n$ & $n-2$ \\
\texttt{EA}  & $n-1$ & $n$ \\
\texttt{DEA} & $n-2$ & $n$ \\
        \hline\hline
    \end{tabular}
%\vspace{-0.4cm}
\end{table}

\subsection{Spin-Orbitals, Spin-Tracing, and Orbital Space Specification }

\label{SEC:SPIN_TRACING}

\pq automatically normal-orders each operator product once they are specified using the  \texttt{add\_operator\_product}, etc. functions. If the vacuum state is the Fermi vacuum, then only fully-contracted terms are retained after this step. If normal order is defined with respect to the true vacuum state, however, \pq will retain all normal-ordered strings. In either case, the resulting list of terms can be simplified by calling
\begin{lstlisting}
pq.simplify()
\end{lstlisting}
which compares the terms to identify the ones that cancel or can be combined based on the antisymmetry properties of the integrals and amplitudes arising in the operators in Table \ref{tab:operators}.

The normal-ordered strings can be extracted from the \pq helper object via the function \texttt{strings}. If normal order is defined with respect to the Fermi vacuum, then this function returns a list of all of the fully-contracted terms. If normal order is defined with respect to the true vacuum, then the list contains all of the normal-ordered terms. In either case, these terms are formatted as lists of strings. If passed a dictionary of spin labels for any non-summed labels (or an empty dictionary if the expression does not involve any non-summed labels), then \pq blocks the terms by spin and only returns those terms that are non-zero based on spin symmetry. 

Consider the singles residual equation in CCSD
\begin{equation}
\label{EQN:SINGLES_RESIDUAL}
    0 = \langle \Phi_0 | \hat{a}^\dagger_i \hat{a}_a \text{exp}\left ( -\hat{T}_1 - \hat{T}_2 \right )\hat{H} \text{exp}\left ( \hat{T}_1 + \hat{T}_2 \right ) | \Phi_0 \rangle
\end{equation}
with $\hat{H} = \hat{f} + \hat{v}$. The following code will output the fully-contracted strings corresponding to this expression (where all orbital labels correspond to spin-orbital labels), as well as a set of spin-blocked terms specific to the case where the orbitals indexed by labels $i$ and $a$ have $\alpha$-spin symmetry.
\begin{lstlisting}
import pdaggerq

pq = pdaggerq.pq_helper('fermi')
pq.set_left_operators([['a*(i)', 'a(a)']])
pq.add_st_operator(1.0,['f'],['t1','t2'])
pq.add_st_operator(1.0,['v'],['t1','t2'])
pq.simplify()

print("# spin-orbital terms")
terms = pq.strings()
for my_term in terms:
    print(my_term)

print("# terms blocked by spin")
spins = {
    'i' : 'a',
    'a' : 'a'
}
terms = pq.strings(spin_labels = spins)
for my_term in terms:
    print(my_term)

pq.clear()
\end{lstlisting}
The corresponding output would be
\begin{lstlisting}
# spin-orbital terms
['+1.00', 'f(a,i)']
['-1.00', 'f(j,i)', 't1(a,j)']
['+1.00', 'f(a,b)', 't1(b,i)']
['-1.00', 'f(j,b)', 't2(b,a,i,j)']
['-1.00', 'f(j,b)', 't1(a,j)', 't1(b,i)']
['+1.00', '<j,a||b,i>', 't1(b,j)']
['-0.50', '<k,j||b,i>', 't2(b,a,k,j)']
['-0.50', '<j,a||b,c>', 't2(b,c,i,j)']
['+1.00', '<k,j||b,c>', 't2(c,a,i,k)', 't1(b,j)']
['+0.50', '<k,j||b,c>', 't2(c,a,k,j)', 't1(b,i)']
['+0.50', '<k,j||b,c>', 't1(a,j)', 't2(b,c,i,k)']
['+1.00', '<k,j||b,i>', 't1(a,k)', 't1(b,j)']
['+1.00', '<j,a||b,c>', 't1(b,j)', 't1(c,i)']
['+1.00', '<k,j||b,c>', 't1(a,k)', 't1(b,j)', 't1(c,i)']
# terms blocked by spin
['+1.00', 'f_aa(a,i)']
['-1.00', 'f_aa(j,i)', 't1_aa(a,j)']
['+1.00', 'f_aa(a,b)', 't1_aa(b,i)']
['-1.00', 'f_aa(j,b)', 't2_aaaa(b,a,i,j)']
['+1.00', 'f_bb(j,b)', 't2_abab(a,b,i,j)']
['-1.00', 'f_aa(j,b)', 't1_aa(a,j)', 't1_aa(b,i)']
['+1.00', '<j,a||b,i>_aaaa', 't1_aa(b,j)']
['+1.00', '<a,j||i,b>_abab', 't1_bb(b,j)']
['-0.50', '<k,j||b,i>_aaaa', 't2_aaaa(b,a,k,j)']
['-0.50', '<k,j||i,b>_abab', 't2_abab(a,b,k,j)']
['-0.50', '<j,k||i,b>_abab', 't2_abab(a,b,j,k)']
['-0.50', '<j,a||b,c>_aaaa', 't2_aaaa(b,c,i,j)']
['+0.50', '<a,j||b,c>_abab', 't2_abab(b,c,i,j)']
['+0.50', '<a,j||c,b>_abab', 't2_abab(c,b,i,j)']
...
\end{lstlisting}
Here, the characters \texttt{a} and \texttt{b} that follow the underscores refer to $\alpha$- and $\beta$-spin, respectively.  Note that we have also introduced the \texttt{clear} function, which clears the list of strings from the \texttt{pq\_helper} object so it could be used again ({\em e.g.} to derive the doubles residual equations, etc.).

\pq also provides support for active-space methods in the style of the CCSDt, CCSDtq, approaches\cite{Bartlett99_6103,Piecuch10_2987} or the CVS approximation.\cite{Dreuw18_7208} Equations for such methods can be obtained by passing a dictionary of label ranges that specifies orbital spaces over which the amplitudes are defined. For both occupied and virtual orbitals, three spaces are defined: \texttt{act}, \texttt{ext}, or \texttt{all}, which refer to active orbitals, external (inactive) orbitals, or the full orbital space. Let us consider the same CCSD singles residual example, but instead of blocking the orbitals by spin, we can block the orbitals by space. As an example, let us restrict \texttt{t2} such that it accounts for at most only one excitation to the external virtual space. The dictionary in the code snippet below achieves this aim, while also indicating that we desire the singles residual equations for the external block of the occupied orbitals and the active block of the virtual orbitals.
\begin{lstlisting}
ranges = {
    't2' : ['all', 'act', 'all', 'all'],
    't1' : ['all', 'all'],
    'a' : ['act'],
    'i' : ['ext']
}
terms = pq.strings(label_ranges=ranges)
for my_term in terms:
    print(my_term)
\end{lstlisting}
The order in which the orbital spaces are specified for \texttt{t2} and \texttt{t1} coincide with the order in which the labels are printed when outputting the fully-contracted strings, {\em e.g.}, for \texttt{t2}, they are ordered as virtual/virtual/occupied/occupied. The corresponding output would be
\begin{lstlisting}
['+1.00', 'f_10(a,i)']
['-1.00', 'f_10(j,i)', 't1_11(a,j)']
['-1.00', 'f_00(j,i)', 't1_10(a,j)']
['+1.00', 'f_11(a,b)', 't1_10(b,i)']
['+1.00', 'f_10(a,b)', 't1_00(b,i)']
['+1.00', 'f_11(j,b)', 't2_1110(b,a,j,i)']
['-1.00', 'f_01(j,b)', 't2_1100(b,a,i,j)']
['-1.00', 'f_10(j,b)', 't2_1010(a,b,j,i)']
['+1.00', 'f_00(j,b)', 't2_1000(a,b,i,j)']
['-1.00', 'f_11(j,b)', 't1_11(a,j)', 't1_10(b,i)']
['-1.00', 'f_10(j,b)', 't1_11(a,j)', 't1_00(b,i)']
['-1.00', 'f_01(j,b)', 't1_10(a,j)', 't1_10(b,i)']
['-1.00', 'f_00(j,b)', 't1_10(a,j)', 't1_00(b,i)']
...
\end{lstlisting}
Here, the characters \texttt{0} and \texttt{1} that follow the underscores refer to external and active orbital spaces respectively. Note that \pq does not currently support simultaneous blocking by spin and by orbital space.

\subsection{Unitary Coupled-Cluster Theory} 
\label{SEC:UCC}

\pq includes functionality for the unitary formulation of CC (UCC). In UCC, the cluster operator, $\hat{T}$, is replaced with its anti-hermitian operator analog, $\hat{\sigma} = \hat{T} - \hat{T}^\dagger$. One can derive equations for UCC or EOM-UCC theory in \pq by specifying 
\begin{lstlisting}
pq.set_unitary_cc(True)
\end{lstlisting}
In this case, an operator product involving \texttt{tn} or \texttt{tn,m} will actually introduce two terms: one that reflects the definition in Table \ref{tab:operators} and one that corresponds to the adjoint of this definition, scaled by a minus sign. The following complications arise when the user requests an anti-hermitian cluster operator. First, the similarity transformation function introduced in Sec.~\ref{SEC:OPERATOR_PRODUCTS} (\texttt{add\_st\_operator}) minimizes computational effort by assuming that the operators that define the transformation ($\hat{c}$ and $\hat{d}$ in Eq.~\ref{EQN:BCH}) commute, but the cluster operators do not commute in UCC. As mentioned above, an optional flag can be passed to this function to indicate that the operators do not actually commute (\texttt{do\_operators\_commute = False}), in which case this assumption is lifted. Second, the \texttt{add\_st\_operator} function assumes that the BCH expansion truncates after four nested commutators, which is not the case for UCC. As such, it is not recommended that users interested in deriving UCC equations use this function. Rather, one can proceed by defining the similarity transformation in the following ways.

Historically, many implementations of UCC have used truncation schemes for the BCH expansion of the similarity-transformed Hamiltonian that are based on perturbation theory arguments.\cite{Noga89_133,Bartlett89_359} As an example, let us consider the UCC3 method, which is an approximation to UCC with single and double excitations where the energy expression is correct to third-order in perturbation theory, and the residual equations are correct to second-order in perturbation theory. To obtain programmable expressions for UCC3 in \pq, the user should use the \texttt{add\_opperator},  \texttt{add\_commutator}, etc.~functions directly to build up an appropriate approximation to the similarity-transformed Hamiltonian. Consider the singles residual equation (Eq.~\ref{EQN:SINGLES_RESIDUAL}), generalized for the UCC3 case:
\begin{equation}
    0 = \langle \Phi_0 | \hat{a}^\dagger_i \hat{a}_a \text{exp}( - \hat{\sigma}  )\hat{H} \text{exp} ( \hat{\sigma} ) | \Phi_0 \rangle
\end{equation}
with $\hat{\sigma} = \hat{\sigma}_1 + \hat{\sigma}_2$ and $\hat{\sigma}_n = \hat{T}_n - \hat{T}_n^\dagger$.  The following code will define and bring to normal order all of the terms that arise in this equation, up to second-order in perturbation theory. Recall that, assuming a Hartree-Fock reference configuration, the Fock operator (\texttt{f}) is a zeroth-order quantity, the fluctuation potential operator (\texttt{v}) and doubles amplitudes (\texttt{t2}) are first-order quantities, and the singles amplitudes (\texttt{t1}) arise at second order. Thus, we have
\begin{lstlisting}
    import pdaggerq

    pq = pdaggerq.pq_helper('fermi')
    pq.set_left_operators([['a*(i)', 'a(a)']])
    pq.set_unitary_cc(True)

    # 0th order
    pq.add_operator_product(1.0, ['f'])

    # 1st order
    pq.add_operator_product(1.0, ['v'])
    pq.add_commutator(1.0, ['f'], ['t2'])


    # 2nd order
    pq.add_commutator(1.0, ['f'], ['t1'])
    pq.add_commutator(1.0, ['v'], ['t2'])
    pq.add_double_commutator(0.5, ['f'], ['t2'], ['t2'])

    pq.simplify()
\end{lstlisting}

Several alternatives to perturbation-theory-based truncation of the UCC $\bar{H}$ have been proposed, including schemes that give the exact energy for a specific number of electrons\cite{Bartlett06_3393} or truncate the so-called Bernoulli expansion of the similarity-transformed Hamiltonian at a specific commutator rank.\cite{Mukherjee18_244110, Cheng22_2281, Cheng21_174102} The \pq package has built-in support for the Bernoulli representation of $\bar{H}$ up to sixth order, {\em i.e.}, 
\begin{align}
    \text{exp} ( -\hat{\sigma} )\hat{H} \text{exp}\left ( \hat{\sigma} \right ) = \bar{H}^0 + \bar{H}^1 + \bar{H}^2 + ... + \bar{H}^6 % + ... %+ \bar{H}^5 + \bar{H}^6
\end{align}
with
\begin{align}
\label{EQN:H0}
        \bar{H}^0 &= f + v \\
        \label{EQN:H1}
        \bar{H}^1 &= [f, \hat{\sigma}] + \frac{1}{2} [v, \hat{\sigma}] + \frac{1}{2} [v_R, \hat{\sigma}] \\
        \label{EQN:H2}
        \bar{H}^2 &= \frac{1}{12}[[v_N, \hat{\sigma}], \hat{\sigma}] + \frac{1}{4}[[v, \hat{\sigma}]_R, \hat{\sigma}] + \frac{1}{4}[[v_R, \hat{\sigma}]_R, \hat{\sigma}] \\
        ... \nonumber
\end{align}
Definitions of $\bar{H}^3$ and $\bar{H}^4$, as well as general recipes for constructing higher-order terms, can be found in Ref.~\citenum{Mukherjee18_244110}. 
In Eqs.~\ref{EQN:H1} and \ref{EQN:H2}, the subscripts $N$ and $R$ refer to the pure excitation / de-excitation parts (up to a specific [de-]excitation order) and the remainder of the operator, respectively. Note that, in this expansion, the Fock operator does not appear in commutators of higher rank than one. 

The Bernoulli expansion of the similarity-transformed fluctuation potential (up to sixth order) can be accessed via the \texttt{add\_bernouli\_operator} function. As an example, the following code will generate equations corresponding to the singles residual for the quadratic UCC with single and double excitations method (qUCCSD),\cite{Cheng21_174102} which includes up to triple commutators in the energy expression and double commutators in the amplitude equations
\begin{lstlisting}
    import pdaggerq

    pq = pdaggerq.pq_helper('fermi')
    
    pq.set_unitary_cc(True)
    pq.set_bernoulli_excitation_level(2)

    pq.set_left_operators([['a(i)*', 'a(a)']])
    pq.add_operator_product(1.0, ['f'])
    pq.add_commutator(1.0, ['f'], ['t1'])
    pq.add_commutator(1.0, ['f'], ['t2'])

    pq.add_bernoulli_operator(1.0,['v'],['t1','t2'], 2)

    pq.simplify()
\end{lstlisting}
Here, the order of the Bernoulli expansion is specified as an input argument to the \texttt{add\_bernoulli\_operator} function. Note also that the \texttt{set\_bernoulli\_excitation\_level} function defines the maximum excitation level at which a pure excitation or de-excitation term will belong to the "N" part of an operator (the default value is 2). 
As an alternative to the \texttt{add\_bernoulli\_operator} function, one may define the same equations via calls to the \texttt{add\_commutator} and \texttt{add\_double\_commutator} and manual specification of the operator portions. For example, the following code would correspond to one of the double commutators that appears in Eq.~\ref{EQN:H2}, $\frac{1}{4}[[v_R, \hat{\sigma}]_R, \hat{\sigma}]$
\begin{lstlisting}
    v = 'v{R,R,A}'
    t1_ARA = 't1{A,R,A}'
    t1_AAA = 't1{A,R,A}'
    t2_ARA = 't2{A,R,A}'
    t2_AAA = 't2{A,R,A}'
    pq.add_double_commutator(0.25,[v],[t1_ARA],[t1_AAA])
    pq.add_double_commutator(0.25,[v],[t1_ARA],[t2_AAA])
    pq.add_double_commutator(0.25,[v],[t2_ARA],[t1_AAA])
    pq.add_double_commutator(0.25,[v],[t2_ARA],[t2_AAA])
    
    pq.simplify()
\end{lstlisting}
Here, the label \texttt{A} refers to "all" of the operator (the combined $N$ and $R$ parts). The order of these operator portion designations corresponds to the placement of the relevant subscripts in the double commutator expression. Using this manual specification, a user could define the Bernoulli representation of $\bar{H}$ up to arbitrary order. 

\subsection{The True Vacuum and Reduced Density Matrices}

In this section, we consider a use case involving reduced density matrices (RDMs) for which it is most convenient to define normal order with respect to the true vacuum. The following code evaluates the orbital gradient
\begin{align}
\label{eqn:orbital_gradient}
g_{tu} = \left \langle \Psi \left | \left [ \hat{a}^\dagger_t \hat{a}_u -  \hat{a}^\dagger_u \hat{a}_t, \hat{H} \right ] \right | \Psi \right \rangle
\end{align}
where $|\Psi\rangle$ is an $N$-electron state, and $\hat{H} = \hat{h} + \frac{1}{4} \hat{g} $, where $\hat{h}$ and $\hat{g}$ are one-electron and antisymmetrized two-electron operators that can be represented in \pq with the operators \texttt{h} and \texttt{g} in Table \ref{tab:operators}, respectively. The following code snippet will bring the operators on the right-hand side of Eq.~\ref{eqn:orbital_gradient} to normal order with respect to the true vacuum state.
\begin{lstlisting}
import pdaggerq

pq = pdaggerq.pq_helper('true')

print("# [t* u - u* t, H]")
pq.add_commutator( 1.0, ['a*(t)', 'a(u)'], ['h'])
pq.add_commutator(-1.0, ['a*(u)', 'a(t)'], ['h'])
pq.add_commutator( 0.25, ['a*(t)', 'a(u)'], ['g'])
pq.add_commutator(-0.25, ['a*(u)', 'a(t)'], ['g'])
pq.simplify()

terms = pq.strings()
for my_term in terms:
    print(my_term)

\end{lstlisting}
The corresponding output would be
\begin{lstlisting}
# [t* u - u* t, H]
['+1.00', 'a*(t)', 'a(p)', 'h(u,p)']
['-1.00', 'a*(p)', 'a(u)', 'h(p,t)']
['-1.00', 'a*(u)', 'a(p)', 'h(t,p)']
['+1.00', 'a*(p)', 'a(t)', 'h(p,u)']
['-0.50', 'a*(p)', 'a*(t)', 'a(q)', 'a(r)', 'g(u,p,r,q)']
['-0.50', 'a*(p)', 'a*(q)', 'a(r)', 'a(u)', 'g(p,q,t,r)']
['+0.50', 'a*(p)', 'a*(u)', 'a(q)', 'a(r)', 'g(t,p,r,q)']
['+0.50', 'a*(p)', 'a*(q)', 'a(r)', 'a(t)', 'g(p,q,u,r)']
\end{lstlisting}
The expectation value of these operators with respect to the $N$-electron state, $|\Psi\rangle$, should be expressible in terms of the elements of the one-electron RDM (1RDM) and the two-electron RDMs (2RDM). Such expressions could have been obtained if we had set
\begin{lstlisting}
pq.set_use_rdms(True)
\end{lstlisting}
at the beginning of that code snippet. In that case, the output would have been

\pagebreak

\begin{lstlisting}
# [t* u - u* t, H]
['+1.00', 'h(u,p)', 'D1(t,p)']
['-1.00', 'h(p,t)', 'D1(p,u)']
['-1.00', 'h(t,p)', 'D1(u,p)']
['+1.00', 'h(p,u)', 'D1(p,t)']
['-0.50', 'g(u,p,r,q)', 'D2(p,t,r,q)']
['-0.50', 'g(p,q,t,r)', 'D2(p,q,u,r)']
['+0.50', 'g(t,p,r,q)', 'D2(p,u,r,q)']
['+0.50', 'g(p,q,u,r)', 'D2(p,q,t,r)']
\end{lstlisting}

\noindent where $\texttt{D1}$ and $\texttt{D2}$ represent the 1RDM and 2RDM, respectively. Some RDM theories make use of the concept of the cumulant decomposition of the 2RDM or higher-order RDMs, where the cumulant or fully connected part of the RDM is discarded in order to close or simplify equations.\cite{Mazziotti99_419,Mazziotti07_104104} As an example, the cumulant decomposition of the 2RDM is
\begin{align}
    {}^2D^{pq}_{rs} = {}^1D^p_r {}^1D^q_s - {}^1D^p_s {}^1D^q_r + {}^2\Delta^{pq}_{rs},
\end{align}
where ${}^1D^p_r$, ${}^2D^{pq}_{rs}$, and ${}^2\Delta^{pq}_{rs}$ represent elements of the 1RDM, 2RDM, and cumulant 2RDM, respectively. In Hartree-Fock theory, the cumulant part of the 2RDM is zero. The orbital gradient for Hartree-Fock could have been obtained by specifying a list of cumulant RDMs that could be ignored when outputting equations involving the RDMs, {\em i.e.},
\begin{lstlisting}
pq.set_use_rdms(True, ignore_cumulant = [2])
\end{lstlisting}
In this case, the resulting output would be
\begin{lstlisting}
# [t* u - u* t, H]
['+1.00', 'h(u,p)', 'D1(t,p)']
['-1.00', 'h(p,t)', 'D1(p,u)']
['-1.00', 'h(t,p)', 'D1(u,p)']
['+1.00', 'h(p,u)', 'D1(p,t)']
['-0.50', 'g(u,p,r,q)', 'D1(p,r)', 'D1(t,q)']
['+0.50', 'g(u,p,r,q)', 'D1(p,q)', 'D1(t,r)']
['-0.50', 'g(p,q,t,r)', 'D1(p,u)', 'D1(q,r)']
['+0.50', 'g(p,q,t,r)', 'D1(p,r)', 'D1(q,u)']
['+0.50', 'g(t,p,r,q)', 'D1(p,r)', 'D1(u,q)']
['-0.50', 'g(t,p,r,q)', 'D1(p,q)', 'D1(u,r)']
['+0.50', 'g(p,q,u,r)', 'D1(p,t)', 'D1(q,r)']
['-0.50', 'g(p,q,u,r)', 'D1(p,r)', 'D1(q,t)']
\end{lstlisting}
In \pq, the \texttt{ignore\_cumulant} flag can be used to approximate the 2RDM or three-particle RDM in terms of lower-order RDMs.

An interesting aspect of these RDM capabilities is that they are necessary ingredients for the description of more general vacua that appear in active-space based multireference theories such as complete active space CI. In such a case, the normal-order engine for the Fermi vacuum could be generalized to consider active orbitals in addition to the restricted occupied and virtual orbitals that define the Fermi vacuum. The multireference vacuum engine should apply Fermi vacuum normal rules to the restricted occupied and virtual orbitals and true vacuum normal order rules to the active orbitals. When combined with the new RDM functionality described above, we would then have everything needed for generating equations and code for active-space-based multireference theories. As such, one of the near term goals for future developments in \pq is exactly this generalization.

\subsection{Analysis of Configuration Interaction and Coupled-Cluster Wave Functions}

Before moving on, we provide one last example showing how \pq can elucidate the relationship between CI and CC amplitudes. Recall that the CI and CC wave functions can be written as $|\Psi_0^\mathrm{CI}\rangle = (\mathbf{1} + \hat{C}) |\Phi_0\rangle$ and $|\Psi_0^\mathrm{CC}\rangle = \text{exp}(\hat{T}) |\Phi_0\rangle$, respectively, with the intermediate normalization condition $\langle\Phi_0|\Psi_0\rangle = 1$. At the full CI/CC limit, the many-body components of $\hat{C}$ and $\hat{T}$ satisfy\cite{Cizek66_4256,Sroubkova69_149,Shavitt72_50}
\begin{align}
\label{eqn:ci_cc_mapping}
    \hat{C}_1 & {}= \hat{T}_1, \\
    \hat{C}_2 & {}= \hat{T}_2 + \frac{1}{2}\hat{T}_1^2, \\
    \label{EQN:C3_FROM_T}
    \hat{C}_3 & {}= \hat{T}_3 + \hat{T}_2\hat{T}_1 + \frac{1}{6}\hat{T}_1^3, \\
    \hat{C}_4 & {}= \hat{T}_4 + \frac{1}{2}\hat{T}_2^2 + \hat{T}_3\hat{T}_1 + \frac{1}{2}\hat{T}_2\hat{T}_1^2 + \frac{1}{24}\hat{T}_1^4,
\end{align}
and so on. While the relationship between the lower-order coefficients and amplitudes are straightforward to derive, their higher-order counterparts can be tricky due to disconnected terms that require careful handling of sign changes due to permutations. The following code translates cluster amplitudes into the corresponding $\hat{C}_3$ coefficients following the  relationship in Eq.~\ref{EQN:C3_FROM_T}.
\begin{lstlisting}
import pdaggerq

pq = pdaggerq.pq_helper("fermi")

# define exp(T) as Taylor series
# T = T1 + T2 + T3 + T4
T = ['t1', 't2', 't3', 't4']
eT = []
# order = 0
eT.append([1.0, ['1']])
# order = 1 -> appears in C1 and above
for my_T in T :
    eT.append([1.0, [my_T]])
# order = 2 -> appears in C2 and above
for my_T1 in T :
    for my_T2 in T :
        eT.append([0.5, [my_T1, my_T2]])
# order = 3 -> appears in C3 and above
for my_T1 in T :
    for my_T2 in T :
        for my_T3 in T :
            eT.append([1.0 / 6.0, [my_T1, my_T2, my_T3]])
                
# c(abc,ijk) = <0|i* j* k* c b a e(T)|0>
pq.set_left_operators([['a*(i)','a*(j)','a*(k)','a(c)','a(b)','a(a)']])
pq.set_right_operators([['1']])

for term in eT:
    pq.add_operator_product(term[0], term[1])

pq.simplify()

terms = pq.strings()
for my_term in terms:
    print(my_term)
\end{lstlisting}
The output of the above code sample contains the properly antisymmetrized $\hat{C}_3$ coefficients with the expected connected and disconnected contributions:

\pagebreak
\begin{lstlisting}
['+1.00', 't3(a,b,c,i,j,k)']
['+1.00', 'P(j,k)', 'P(a,b)', 't1(a,k)', 't2(b,c,i,j)']
['+1.00', 'P(a,b)', 't1(a,i)', 't2(b,c,j,k)']
['+1.00', 'P(j,k)', 't2(a,b,i,j)', 't1(c,k)']
['+1.00', 't2(a,b,j,k)', 't1(c,i)']
['-1.00', 'P(i,j)', 't1(a,k)', 't1(b,j)', 't1(c,i)']
['+1.00', 'P(i,k)', 't1(a,j)', 't1(b,k)', 't1(c,i)']
['-1.00', 'P(j,k)', 't1(a,i)', 't1(b,k)', 't1(c,j)']
\end{lstlisting}

\section{Code Generation}

\label{sec:code}

The current version of \pq has two modules for generating usable computer code corresponding to normal-ordered expressions such as those discussed in the preceding sections. The \texttt{parser} module was part of the original release of \pq and is capable of generating Python code with limited floating-point optimization capabilities. More recently, we have developed a second module, \texttt{pq-graph}, which has more robust floating-point and memory optimization protocols and can generate either Python or C++ code. 

\subsection{The \texttt{parser} Module}

The \texttt{parser} module translates output of the \texttt{strings} function into Python code that carries out tensor contractions using calls to \np~ \texttt{einsum}. The floating-point cost for individual tensor contractions can automatically be optimized via \texttt{einsum}'s \texttt{optimize=optimal} flag.\cite{Gray18_753} Appropriate limits on the summation labels ({\em e.g.}, occupied, virtual, or general orbitals) is enforced using array slicing.

The \texttt{parser} module has been updated to reflect changes in the equation generation capabilities discussed above. First, the \texttt{parser} model recognizes labels that are added when the equations are blocked according to spin symmetry or orbital space. Related, the array slicing has been generalized to account for different occupied and virtual spaces corresponding to different spin symmetries or spatial orbital spaces. Third, the \texttt{parser} model has been generalized to recognize additional tensor quantities. Examples include the 1-, 2-, 3-, and 4-electron RDMs (\texttt{D1}, \texttt{D2}, \texttt{D3}, and \texttt{D4}, respectively) and photon / mixed electron-photon quantities ({\em e.g.}, \texttt{w0}, \texttt{d+}, \texttt{d-}, \texttt{tn,m}, etc.). 

As an example, let us consider the spin-traced CCSD singles residual example from Sec.~\ref{SEC:SPIN_TRACING}. The following code generates the relevant \texttt{einsum} expressions
\begin{lstlisting}
import pdaggerq

pq = pdaggerq.pq_helper('fermi')
pq.set_left_operators([['a*(i)', 'a(a)']])
pq.add_st_operator(1.0,['f'],['t1','t2'])
pq.add_st_operator(1.0,['v'],['t1','t2'])
pq.simplify()

spins = {
    'i' : 'a',
    'a' : 'a'
}
terms = pq.strings(spin_labels = spins)

from pdaggerq.parser import contracted_strings_to_tensor_terms

tensor_terms = contracted_strings_to_tensor_terms(terms)

for my_term in tensor_terms:
    einsum_terms = my_term.einsum_string(update_val='r1_aa', output_variables=('a', 'i'))
    print("%s" % (einsum_terms))
\end{lstlisting}
The output of this code is
\begin{lstlisting}
r1_aa +=  1.00 * einsum('ai->ai', f_aa[va, oa])
r1_aa += -1.00 * einsum('ji,aj->ai', f_aa[oa, oa], t1_aa)
r1_aa +=  1.00 * einsum('ab,bi->ai', f_aa[va, va], t1_aa)
r1_aa += -1.00 * einsum('jb,baij->ai', f_aa[oa, va], t2_aaaa)
r1_aa +=  1.00 * einsum('jb,abij->ai', f_bb[ob, vb], t2_abab)
r1_aa += -1.00 * einsum('jb,aj,bi->ai', f_aa[oa, va], t1_aa, t1_aa, optimize=['einsum_path', (0, 1), (0, 1)])
...
\end{lstlisting}
Note that slices corresponding to the different spin cases arise for the occupied orbitals (\texttt{oa} and \texttt{ob}) and virtual orbitals (\texttt{va} and \texttt{vb}); it is left to the user to define these array slices, as well as any other required tensors ({\em e.g.}, \texttt{f\_aa}, etc.) within an actual code. Note also that the last term passes the \texttt{optimize} flag to \texttt{einsum}, which 
performs an exhaustive search of tensor contraction orderings to give the lowest scaling. This single-term analysis represents the extent of the floating-point optimization capabilities of the \texttt{parser} module. Additional optimization protocols have been developed within the \texttt{pq\_graph} module, which is described in the next section.

\subsection{The \pqgraph Module}
While the \texttt{parser} module translates the string representations of the tensor contractions obtained from the \texttt{strings} function into calls to \np~ \texttt{einsum}, the \pqgraph module works directly with the internal representation of the normal-ordered strings within \pq. It represents each tensor contraction with a directed acyclic graph (DAG), which allows for flexible and efficient manipulation of quantum chemistry expressions. Each vertex in the DAG corresponds to a tensor, and the edges denote the indices within a contraction.

A contraction is represented as a binary tree, internally referred to as a \texttt{Linkage}, where each linkage points to two children --- a left and right vertex --- representing the tensors being contracted. Importantly, since any contraction may be replaced by an intermediate tensor, a \texttt{Linkage} is also treated as a \texttt{Vertex}. In other words, any sequence of nested contractions is treated as a single composite vertex; even deeply nested expressions can be easily identified and substituted just as any other tensor. The recursive structure simplifies traversal and optimizations across entire equations. Consequently, the \pqgraph module can efficiently identify optimal contraction paths, detect shared subcontractions across different terms, and fuse equivalent intermediates into a single computational expression.

\subsubsection{Single-Term Optimization}

Single-term optimization exhaustively rearranges the contraction order within a single term to reduce computational cost. It aims to find a sequence of binary contractions that yields the lowest total number of floating-point operations (FLOPs). This type of optimization is similar to that invoked by passing the \texttt{optimize=optimal} to a \np~\texttt{einsum} call. However, \pqgraph extends this capability by exposing full control over permutations and offering user-configurable options to modify how contractions are selected.

The algorithm operates as follows for any given term:
\begin{enumerate}
\item Generate all permutations of binary contraction orders.
\item For each permutation, compute the computational cost with a function that considers the rank of the intermediate tensors (e.g., $O(o^{x}v^{y})$ for $x$ occupied and $y$ virtual orbital indices), as well as the memory cost of the intermediates.
\item Select the permutation with the lowest total cost as the optimal contraction order.
\end{enumerate}

The cost function balances FLOP minimization and intermediate tensor sizes. When multiple contraction orders yield the same FLOP count, the algorithm prefers those with lower memory requirements when generating intermediate tensors. Additionally, users can enforce maximum intermediate storage constraints, making optimization aware of practical hardware limits. This flexibility proves valuable in high-performance settings, where memory bottlenecks may outweigh FLOP-based metrics.

\subsubsection{Subexpression Elimination}

Subexpression elimination identifies similar tensor contractions, \emph{i.e.} subexpressions, that appear across multiple terms. Instead of re-evaluating the same repeated contractions, the subexpressions can be replaced with reusable intermediate tensors. Eliminating redundant evaluations dramatically reduces the total number of operations and simplifies the final set of equations. The algorithm for subexpression elimination in \pqgraph proceeds through the following steps:

\begin{enumerate}
\item Analyze the DAG structures of all terms to enumerate unique subexpression candidates.
\item For each subexpression:
\begin{enumerate}
\item Generate a temporary intermediate to store the subexpression result.
\item Attempt to substitute every occurrence of this subexpression in the original terms with the new intermediate. Replacements are accepted only when the resulting FLOP count for each term remains the same or improves.
\item Calculate the overall cost of this replacement across all affected terms and record it.
\item Revert the replacements to test the next candidate independently.
\end{enumerate}
\item Rank the subexpression candidates by their total cost savings.
\item Apply the substitutions in order of descending cost benefit. Note that some candidates may conflict with one another, making it impossible to apply all optimizations simultaneously.
\item Remove redundant or shadowed subexpressions, finalize the intermediate naming scheme, and update the DAGs.
\item Repeat the process until no more replacements yield improvements.
\end{enumerate}

Subexpression matching uses the \texttt{Linkage} class to determine equality based on tensor identity, matching contraction indices, and recursive structure.  This allows rapid grouping of similar contractions across terms, retaining only unique subexpressions as candidates to test for elimination.

\subsubsection{Fusion}

Fusion improves the efficiency of code generation by merging intermediate tensors that are used in similar ways across multiple terms. Instead of evaluating several similar contractions independently, fusion allows their common elements to be grouped together. For example, the pair of contractions $a^p_r b^r_q + c^p_r b^r_q$ can be rewritten as a fused intermediate $d^p_r = a^p_r + c^p_r$, followed by a single contraction $d^p_r b^r_q$.

The fusion algorithm proceeds as follows:
\begin{enumerate}
\item Construct a map linking each intermediate to the terms that use it.
\item Group intermediates by shape and contraction structure.
\item For each group, identify compatible intermediates by comparing the operations that use them.
\item Fuse compatible intermediates by summing them, updating the terms that depend on them, and removing the originals.
\end{enumerate}
Fusion can significantly reduce the number of terms and the computational complexity of the final code. It also reduces the number of temporary tensors, and consequently the memory requirements for a calculation.

\subsubsection{Equation Optimization and Analysis}

As a practical example, the following code uses \pqgraph to generate C++ code corresponding to the spin-orbital representation of the CCSD doubles residual equations:
\begin{align}
\label{EQN:DOUBLES_RESIDUAL}
    0 = \langle \Phi_0 | \hat{a}^\dagger_i \hat{a}^\dagger_j \hat{a}_b \hat{a}_a \text{exp}( -\hat{T} )\hat{H} \text{exp} ( \hat{T} )| \Phi_0 \rangle
\end{align}
with $\hat{T} = \hat{T}_1 + \hat{T}_2  $.
\begin{lstlisting}
import pdaggerq

pq = pdaggerq.pq_helper('fermi')
pq.set_left_operators([['a*(i)','a*(j)','a(b)','a(a)']])
pq.add_st_operator(1.0, ['f'], ['t1', 't2'])
pq.add_st_operator(1.0, ['v'], ['t1', 't2'])
pq.simplify()

# initialize a pq_graph object
options = {}
graph = pdaggerq.pq_graph(options)

# optimize equations
graph.add(pq, "r2", ['a', 'b', 'i', 'j'])
graph.optimize() 

# print equations
graph.print('c++') 
\end{lstlisting}
The output of this code includes the following information: (i) a list of valid options for initializing the \texttt{pq\_graph} object, (ii) a list of tensors that should be initialized in order for the subsequent C++ code to run properly ({\em e.g.}, electron repulsion integrals, etc.), and (iii) the optimized C++ code corresponding to the CCSD doubles residual equations, where tensor contractions are carried out using the TiledArray or \np~libraries.

The \pqgraph module can also provide a detailed summary of the optimizations applied and the associated computational savings, which can be generated by the following call, after the \texttt{optimize} step
\begin{lstlisting}
graph.analysis()   
\end{lstlisting}
The output of the \texttt{analysis} function includes a breakdown of the FLOP scaling (with respect to the number of occupied [\texttt{o}] or virtual [\texttt{v}] orbitals) for the terms present in the equations:
\begin{lstlisting}

Total FLOP scaling: 
------------------
 Scaling :   I   |   R   |   F   ||  F-I  |  F-R 
-------- : ----- | ----- | ----- || ----- |  ----
    o3v4 :     5 |     0 |     0 ||    -5 |     0 
    o4v3 :     3 |     0 |     0 ||    -3 |     0 
-------- : ----- | ----- | ----- || ----- |  ----
    o2v4 :     9 |     1 |     1 ||    -8 |     0 
    o3v3 :     8 |     5 |     5 ||    -3 |     0 
    o4v2 :     1 |     8 |     7 ||     6 |    -1 
-------- : ----- | ----- | ----- || ----- |  ----
    o1v4 :     8 |     1 |     1 ||    -7 |     0 
    o2v3 :    23 |    10 |     9 ||   -14 |    -1 
    o3v2 :     5 |    25 |    23 ||    18 |    -2 
    o4v1 :     1 |     8 |     6 ||     5 |    -2 
-------- : ----- | ----- | ----- || ----- |  ----
    o1v3 :     0 |     1 |     1 ||     1 |     0 
    o2v2 :    32 |    34 |    34 ||     2 |     0 
    o3v1 :     0 |     1 |     2 ||     2 |     1 
    o4v0 :     0 |     0 |     3 ||     3 |     3 
-------- : ----- | ----- | ----- || ----- |  ----
    o1v2 :     1 |     0 |     0 ||    -1 |     0 
    o2v1 :     0 |     2 |     2 ||     2 |     0 
-------- : ----- | ----- | ----- || ----- |  ----
    o1v1 :     0 |     0 |     1 ||     1 |     1 
-------- : ----- | ----- | ----- || ----- |  ----
   Total :    96 |    96 |    95 ||    -1 |    -1 
\end{lstlisting}
The second column (labeled {\texttt{I}}) indicates the number of terms displaying this scaling given in the first column, before any optimization. The next two columns (labeled {\texttt{R}} and {\texttt{F}}) provide the number of terms with each scaling after the single-term optimization step ({\texttt{R}}) and after the subexpression elimination and fusion steps ({\texttt{F}}). Note that the number of terms of a given scaling do not necessarily decrease at each stage. As an example, fusion will decrease the number of high-scaling contractions at the expense {of increasing lower-scaling operations}. 
Additional examples of how to use the \pqgraph module to generate optimized C++ and Python code are provided on GitHub.\cite{github_pdaggerq}

Lastly, we demonstrate how the spin tracing and \texttt{pq-graph} optimization features in \pq impact computational efficiency in a real example. Specifically, the data in Table \ref{TAB:CCSDT_TIMINGS} correspond to the time required to evaluate the CCSDT residual equations for hydrogen fluoride represented by the cc-pVDZ basis.\cite{Dunning89_1007} The timings in represent the average time per iteration (in seconds) for four Python implementations of automatically generated codes using (i) spin-orbital expressions, (ii) spin-orbital expressions optimized using the \texttt{pq-graph} module, (iii) spin-integrated expressions, and (iv) spin-integrated expressions optimized using the \texttt{pq-graph} module.  The residual equations were incorporated into a modified version of the CCSDT code provided in the examples in the \pq GitHub repository, which obtains reference orbitals and molecular integrals from the \textsc{Psi4} package.\cite{Sherrill20_184108} The implementations are consistent in the sense that, when converging the energy to below 10$^{-10}$ E$_\text{h}$ and the residual equations to 10$^{-10}$, the correlation energy obtained from each implementation differs by less than 10$^{-12}$ E$_\text{h}$. For the spin-orbital implementations, \texttt{pq-graph} optimizations lead to a factor of $\approx$ 2.2 reduction in the time per CCSDT iteration for this system. Spin integration leads to substantially larger improvements. Without \texttt{pq-graph} optimization, the time to evaluate the residual equations using the spin-integrated is nearly 16 times less than that required by the spin-orbital code. $\texttt{pq-graph}$ optimization of the spin-integrated code improves the performance by another factor of $\approx$ 1.5.

\begin{table}[!htpb]
    \centering
    \caption{The average time to evaluate the CCSDT residual equations (seconds) for hydrogen fluoride represented by the cc-pVDZ basis. }
    \label{TAB:CCSDT_TIMINGS}
    \begin{tabular}{lcc}
    \hline\hline
            spin treatment & \texttt{pq-graph} optimization & time \\
        \hline
spin-orbital     &  no & 127.4 \\
spin-orbital     & yes &  58.4 \\
spin-integrated  &  no &   8.2 \\
spin-integrated  & yes &   5.3 \\
        \hline\hline
    \end{tabular}
%\vspace{-0.4cm}
\end{table}

\section{Conclusions}

\label{sec:conclusions}

Over the last few years, the functionality of the \pq package has expanded to cover a large swath of single-reference quantum chemistry methods. On the equation generation side, the current version of the library can produce equations for many flavors of CC and EOM-CC theory, including non-particle-conserving forms of EOM-CC, as well as unitary and cavity QED generalizations thereof. The practical utility of the equation generation engine has also increased with additional support for the specification of electronic spin degrees of freedom and multiple electronic orbital subspaces. 

On the code generation side, the \pqgraph module introduces  optimization techniques tailored for the equations that arise in the many-body quantum chemistry mentioned above.
These techniques, grounded in graph-theoretical principles, are designed to minimize the number of floating-point operations and manage memory requirements effectively. By automating code generation for both Python and C++ workflows, the current version of \pq also caters to a wider audience of developers than the original release of the library.  \\

\noindent {\bf ACKNOWLEDGMENTS:} \\

This material is based upon work supported by the U.S. Department of Energy, Office of Science, Office of Advanced Scientific Computing Research and Office of Basic Energy Sciences, Scientific Discovery through the Advanced Computing (SciDAC) program under Award No. DE-SC0022263 and the National Science Foundation under Grants No. CHE-2100984 and OAC-2103705.\\ 

\bibliography{Journal_Short_Name,main,cc,deprince,rdm,code_generation}

\end{document}